\documentclass[aps,prb,twocolumn,superscriptaddress]{revtex4-1}

\usepackage{graphicx}

\usepackage{amsmath}
\usepackage{amssymb}
\usepackage{bm}
\usepackage{cancel}
\usepackage[normalem]{ulem}
\usepackage{xfrac}
\usepackage[colorlinks=true,linkcolor=blue,citecolor=blue,urlcolor=blue]{hyperref}

\usepackage{textcomp}

\usepackage{color}
\usepackage{xcolor}

\usepackage{hyperref}
\usepackage{cleveref}

\usepackage{subfigure}
\usepackage{placeins}
\usepackage{siunitx}

\newcommand{\bea}{\begin{eqnarray}}
\newcommand{\eea}{\end{eqnarray}}
\newcommand{\be}{\begin{equation}}
\newcommand{\ee}{\end{equation}}
\newcommand{\benn}{\begin{equation*}}
\newcommand{\eenn}{\end{equation*}}

\newcommand{\kB}{k_\text{B}}
\newcommand{\kBT}{k_\text{B}T}

\newcommand{\rmA}{\text{A}}
\newcommand{\rmB}{\text{B}}
\newcommand{\rmC}{\text{C}}
\newcommand{\rmAB}{\text{AB}}
\newcommand{\rmL}{\mathrm{L}}
\newcommand{\rmR}{\mathrm{R}}

\definecolor{light-gray}{gray}{0.9}
\definecolor{gris}{gray}{0.5}

\begin{document}

\title{
\vspace{-2.25cm}
\textnormal{\small  PHYS. REV. B {\bf 98}, 035414 (2018)}\\
\vspace*{0.2cm}
Thermal rectification with interacting electronic channels: Exploiting degeneracy, quantum superpositions and interference}

\author{Alejandro Marcos-Vicioso}

\affiliation{Escuela Polit\'ecnica Superior, Universidad Carlos III de Madrid, 28911 Legan\'es, Spain}

\author{Carmen L\'opez-Jurado} 

\affiliation{Escuela Polit\'ecnica Superior, Universidad Carlos III de Madrid, 28911 Legan\'es, Spain}

\author{Miguel Ruiz-Garcia}

\affiliation{Gregorio Mill\'an Institute for Fluid Dynamics, Nanoscience, and Industrial Mathematics, and Department of Materials Science and Engineering,
Universidad Carlos III de Madrid, 28911 Legan\'es, Spain}

\affiliation{Department of Physics and Astronomy, University of Pennsylvania, Philadelphia, Pennsylvania 19104, USA}

\author{Rafael S\'anchez}

\affiliation{Gregorio Mill\'an Institute for Fluid Dynamics, Nanoscience, and Industrial Mathematics, and Department of Materials Science and Engineering,
Universidad Carlos III de Madrid, 28911 Legan\'es, Spain}

\affiliation{Departamento de F\'isica Te\'orica de la Materia Condensada, Universidad Aut\'onoma de Madrid, 28049 Madrid, Spain}


\begin{abstract}
This paper explores different mechanisms that induce thermal rectification in the nanoscale. The presence of interacting energy  channels combined with simple asymmetries is sufficient for promoting the desired behavior.  
We use simple quantum dot configurations, identifying the basic properties that enhance rectification for each case: the size of a quantum dot state space (which suggests the use of scaled up systems with many interacting channels), tunneling asymmetries due to coherent tunneling in a double quantum dot, or quantum interference in a triangular triple quantum dot. An efficient and tunable thermal diode is proposed using a channel capacitively coupled to a mesoscopic switch.

\end{abstract}
\vspace{-0.3cm}
\maketitle

\section{Introduction}
\label{sec:intro}

\vspace{-0.3cm}
Conventional electronics is constantly increasing computational power through the miniaturization of its basic components. The operation of a circuit can be seriously harmed by the heat dissipated in its working components. Gaining control over nanoscopic heat currents is hence vital for improving the operation of electronic devices. 

There are different approaches to alleviate or even take advantage of the dissipated heat in electronic devices. One possibility is to convert it into power by thermoelectric engines~\cite{benenti_fundamental_2017,review}. One can also think of doing useful operations which are only driven by heat~\cite{dubi_heat_2011,li_phononics_2012}, for which one needs to find thermal analogs of electrical components like a transistor or a diode which work at the nanoscale. Research in this direction has been boosted by the recent advances in the detection of mesoscopic heat currents~\cite{martinez_coherent_2014,jezouin_quantum_2013,riha_mode_2015,cui_quantized_2017,dutta_thermal_2017}.

Any diode, including thermal ones, require a spacial asymmetry that affects the current propagation between two terminals~\cite{benenti_from_2016}.  This has led to proposals of thermal rectifiers based on broken mirror symmetry that use a series of systems with different spectral densities. Recent examples include linear lattices~\cite{li_thermal_2001}, superconducting junctions~\cite{ruokola_thermal_2009,martinez_efficient_2013,oettinger_heat_2014,martinez_rectification_2015}, normal-superconducting junctions~\cite{giazotto_thermal_2013},  metallic islands~\cite{ruokola_single_2011}, quantum Hall tunnel barriers~\cite{vannucci_interference_2015}, metal-dielectric interfaces~\cite{ren_heat_2013}, qubits~\cite{schaller_collective_2016,ordonez_quantum_2017}, or resonators~\cite{barzanjeh_manipulating_2018}. Other possiblities include energy-dependent couplings~\cite{segal_single_2008} or the asymmetric coupling to a third reservoir acting as an environment with which the system exchanges energy~\cite{segal_nonlinear_2008,diode,jiang_phonon_2015,guillem,thermal-long}.
 
One can also use the effect of electron-electron interactions. In quantum dot systems with discrete spectral densities, they are responsible for strong nonlinearities~\cite{scheibner_quantum_2008,xue2008thermal,kuo_thermoelectric_2010,fahlvik_nonlinear_2013,sierra_strongly_2014,sierra_nonlinear_2015,
svilans_nonlinear_2016,josefsson_quantum_2017}.
Several quantum dots can be coherently coupled to form different configurations, which enables one to locally control the density of states. For instance, the symmetry of quantum superpositions can be controlled by gate voltages in linearly-coupled double~\cite{vanderwiel_electron_2002,holgerDQD} or triple quantum dots~\cite{hsieh_physics_2012}. More complicated spacial arrangements~\cite{vidan_triple_2004,kotzian_channel_2016,noiri_triangular_2017} introduce different tunneling paths which give rise to quantum interference effects~\cite{entinwohlman,michaelis,poltl_two-particle_2009,sbcpt,donarini_interference_2010}, under the appropriate symmetries~\cite{dani}. 

\begin{figure}[b]
\includegraphics[width=0.85\linewidth]{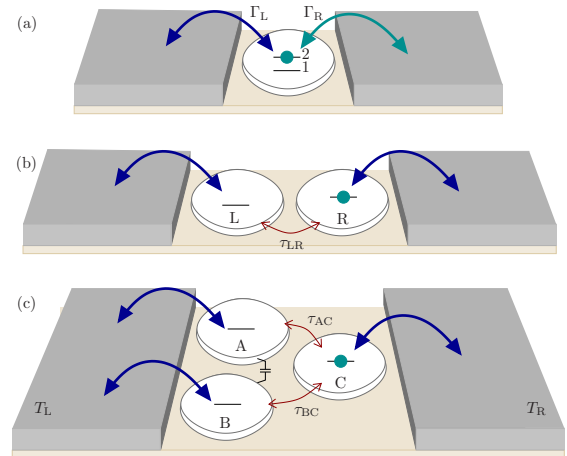}
\caption{Scheme of the different quantum dot configurations discussed in the paper. (a) A single quantum dot is coupled to two terminals with asymmetric tunneling rates $\Gamma_l$. There are two possible states the electrons can tunnel to. (b) A double quantum dot in series. Hybridization of the quantum dot orbitals due to coherent tunneling, $\tau_{\rm LR}$, introduces L-R asymmetric and energy-dependent tunneling rates even if the left and right barriers are identical. (c) A triple quantum dot in a triangular configuration introduces directionality due to the interference of trajectories coming from the left lead. 
}
\label{scheme}
\end{figure}

In this paper we will restrict ourselves to the study of simple configurations of quantum dots, see Fig.~\ref{scheme}. With this minimalistic approach we aim to reduce the number of degrees of freedom helping us to identify the relevant processes.  
In all the considered configurations transport occurs via channels that are correlated via strong Coulomb interactions. We assume the Coulomb blockade regime,
where the system can be occupied by up to one electron. The mechanism which breaks mirror symmetry and enables rectification is different in each case. 

In the simplest case, a single quantum dot is tunnel-coupled to two terminals, the quantum dot levels need to be asymmetrically coupled to the left and right leads, but remarkably no energy-dependent tunneling is needed, cf. Fig.~\ref{scheme}(a). Indeed, we find the surprising result that the multiplicity of the quantum dot states increases the rectification. The left-right asymmetry and energy-dependence of the tunneling rates can be manipulated by controlling the tunneling hybridization in a double quantum dot, cf. Fig.~\ref{scheme}(b), which helps to increase the rectification coefficient. More drastically, a spacially-asymmetric composition with two quantum dots coupled to the left lead and only one coupled to the right can work as a rectifier even if the tunnelig barriers are all identical, cf. Fig.~\ref{scheme}(c). The effect is in this case due to the formation of superpositions of the quantum dots coupled to the left which avoid the occupation of the remaining quantum dot~\cite{amaha_resonance_2012,maria,floris,spintransfer}. They form a blocking channel in parallel to those that support the current. This property can be then applied to simpler configurations with two capacitively coupled quantum dots in parallel~\cite{chan_strongly_2002,hubel_two_2007}, where fluctuations in one quantum dot affect the current through the other one ~\cite{hotspots,holger}. This process is reminiscent of dynamical channel blockade~\cite{belzig,rf} and achieves huge rectification coefficients at configurations with maximal heat currents.

The paper is organized as follows. In Sec.~\ref{sec:model} we present the general formalism which is applied to the different configurations. The effect of the dimensionality of the state space is analyzed in Sec.~\ref{sec:qd}, and asymmetries arising from coherent tunneling are introduced in Sec.~\ref{sec:dqd}. A triple quantum dot where a superposition of states is only coupled to one of the terminals is presented in Sec.~\ref{sec:tqd}, whereas a similar effect is used in a simpler configuration with two quantum dots in Sec.~\ref{sec:qdqd}. Conclusions are discussed in Sec.~\ref{sec:conclusions}.

\vspace{-0.65cm}
\section{Model and equations} 
\label{sec:model}

\vspace{-0.3cm}
Along this work we will show how different configurations of quantum dots can be tuned to exhibit a thermal-diode behavior. In particular, we will consider systems with one, two, and three quantum dots connected to two reservoirs at different temperatures, cf. Fig.~\ref{scheme}. We furthermore assume that the electrostatic charging energy of any of these systems is sufficiently large that the total number of electrons does not exceed one. 

For the systems appearing in Fig. \ref{scheme} we are only interested on the stationary thermal currents. The information of the occupation of every energy level, accounting also for coherences between them, is contained in the density matrix $\rho$. Its evolution follows a master equation
\begin{equation}
\label{eq_lindblad}
\frac{\rm d}{{\rm d}t}{\hat{\rho}}=-\frac{\rm i}{h}[\hat H_{\rm s},\hat\rho] +\sum_{l,X,\alpha}{\cal D}[\hat L_{lX\alpha},\hat\rho],
\end{equation}
where the index $l$=L,R accounts for the left and right leads, and $\alpha=\pm$ refers to tunneling in/out of the system. The first term on the right hand side of Eq. \eqref{eq_lindblad} accounts for the coherent evolution of the isolated system. The second term introduces the tunneling between the system and the reservoirs, which will be specified in terms of the relevant states $X$ of each configuration.  We assume a Born-Markov approximation, valid in the weak tunneling regime, $\Gamma_{lX}\ll\kBT$~\cite{beenakker}, see below. 
The dissipator is given by the usual Lindblad form:
\begin{align}
{\cal D}[\hat L,\hat\rho] = \hat L\hat\rho \hat L^\dagger - \frac{1}{2} \left\{ \hat L^\dagger \hat L,\hat\rho\right\}_+.
\end{align}
In particular,
\begin{align}
\hat L_{lX+}&= \sqrt{\Gamma_{lX}^{+}} |X\rangle \langle 0 |\\
\hat L_{lX-}&= \sqrt{\Gamma_{lX}^-} |0\rangle \langle X |,
\end{align}
where the rates for tunneling in/out are:
$\Gamma_{lX}^+ = \Gamma_{lX} f(E_X{-}\mu_l, T_l)$, and $\Gamma_{lX}^- = \Gamma_{lX}  [1- f(E_X{-}\mu_l, T_l)]$, with the Fermi function $f(E,T)=\left[1+\exp(E/\kBT)\right]^{-1}$ giving the electronic distribution of a lead at temperature $T$. The transparency of barrier $l$, $\Gamma_{lX}$, may depend on the energy level involved in the transition, $X$. We assume a wide band approximation such that any $X$-dependence of the tunneling transparency is only due to the internal spectrum of the quantum dot system. Finally, the equilibrium Fermi energy has been set to zero for convenience. In most cases, we will assume that there is no electric potential applied to the reservoirs, i.e. $\mu_l=0$, except when explicitly mentioned. 

With the stationary solution of the master equation \eqref{eq_lindblad} for the density matrix elements, $\dot\rho_{ij}=0$, we obtain the dc charge and heat currents:
\begin{align}
\label{eq:Il}
I_l &= e\sum_{X}  (\rho_{00} \Gamma_{lX}^+ - \rho_{XX} \Gamma_{lX}^-)\\
\label{eq:Jl}
J_l &= \sum_{X} (E_X-\mu_l) (\rho_{00} \Gamma_{lX}^+ - \rho_{XX} \Gamma_{lX}^-),
\end{align}
respectively. As we are interested in the rectification of heat flows, we need to compute the heat current through the system in response to opposite temperature gradients. 

There are two possible configurations to be considered: (i) short circuit, where $eV=\mu_\rmL-\mu_\rmR=0$, and (ii) open circuit, where a (thermo)voltage develops to the condition $I_l=0$. This paper will mostly focus on case (i), although case (ii) will be discussed for some configurations. In both cases, no power is generated in the system, such that heat is conserved: $J_\rmL+J_\rmR=0$. There is hence no ambiguity in defining the forward and backward responses in a single terminal, e.g. $J^+{=}J_\rmL(T_\rmL{=}T+\Delta T,T_\rmR{=}T)$ and $J^-{=}{-}J_\rmL(T_\rmL{=}T,T_\rmR{=}T{+}\Delta T)$, depending on what terminal has a temperature increase $\Delta T$.

With these, we define the rectification coefficient:
\begin{equation}
R = \left| \frac{J^+ - J^-}{J^+ + J^-} \right|,
\label{eq_R}
\end{equation}
which is bounded between 0 (no rectification) and 1 (an ideal thermal diode).

\vspace{-0.4cm}
\section{Degeneracy in a quantum dot}
\label{sec:qd}

\vspace{-0.3cm}
Let us first consider the case of a single quantum dot~\cite{jens,erdman_thermoelectric_2017}, see Fig.~\ref{scheme}(a). 
It is important to emphasize that mirror symmetry breaking is not sufficient to produce rectification. To show this, it is useful to first explore a simple model with only one available energy state $\varepsilon$ in the quantum dot. This is the case if one can neglect the effect of spin (for instance, if the leads are fully spin-polarized). The coupling of this state to the left and right leads is parametrized by the tunneling rates $\Gamma_{l1}$. The heat current (at the condition of no voltage bias) is given by:
\be
\label{onestate}
J_1(\varepsilon)=\varepsilon\frac{\Gamma_{\rmL1}\Gamma_{\rmR1}}{\Gamma_{\rmL1}{+}\Gamma_{\rmR1}}[f(\varepsilon,T_\rmL)-f(\varepsilon,T_\rmR)],
\ee
as shown in App.~\ref{app:onelevel}.
As the temperature dependence only enters in the difference of Fermi functions, the current is antisymmetric under the change $T_\rmL\leftrightarrow T_\rmR$, leading to $J^+=J^-$, i.e. no rectification. Note also that having additional states does not change the situation provided that they do not interact with each other, as discussed in App.~\ref{app:manyonelevel}. 
\begin{figure}[t]
\includegraphics[width=0.8\linewidth]{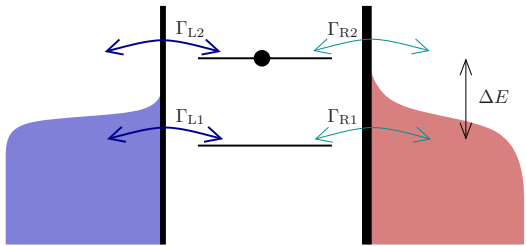}
\caption{Two-level quantum dot coupled to two terminals. Left-right asymmetric and level-dependent tunneling lead to four different tunneling rates, $\Gamma_{li}$. The latest can be due to an energy splitting $\Delta E$ in energy-dependent barriers, or to an additional degree of freedom (e.g. spin). }
\label{QDlevels}
\end{figure}

However, the behavior dramatically changes when considering several channels that are correlated via interactions. We are interested here in the simplest case of two states that exclude each other (due to strong Coulomb blockade), as depicted in Fig.~\ref{QDlevels}. We label them $|X\rangle=|1\rangle,|2\rangle$ for simplicity. In our case, they can correspond to the two possible spin states of the electron that occupies a single-level quantum dot, which can be split by $\Delta E$, e.g. due to an applied magnetic field. We remark here that the interplay of two states (a ground and an excited state) was used to interpret the rectification of a quantum dot in experiments~\cite{scheibner_quantum_2008}.

As there are no internal dynamics in the quantum dot, the first term on the right-hand side of Eq.~\eqref{eq_lindblad} does not contribute to the evolution of the system. Hence, off-diagonal elements of the density matrix are uncoupled from the occupations $\rho_{ii}$ and need not be taken into account. The master equation then simply reads:
\be
\dot \rho_{ii}=\Gamma_{\Sigma i}^+\rho_{00}-\Gamma_{\Sigma i}^-\rho_{ii},\quad i{=}1,2
\label{matrixQD}
\ee
and is complemented with the normalization condition $\rho_{00}+\rho_{11}+\rho_{22}=1$. Here we use the notation $\Gamma_{\Sigma i}^\pm=\sum_l\Gamma_{li}^\pm$, with $l$=L,R.

The steady-state solution of Eq.~\eqref{matrixQD} reads: $\rho_{00}=\Gamma_{\Sigma1}^-\Gamma_{\Sigma2}^-/\Lambda$, $\rho_{11}=\Gamma_{\Sigma1}^+\Gamma_{\Sigma2}^-/\Lambda$, and $\rho_{22}=\Gamma_{\Sigma1}^-\Gamma_{\Sigma2}^+/\Lambda$. The common parameter in the denominator accounts for normalization:
\be
\Lambda=\Gamma_{\Sigma1}^-\Gamma_{\Sigma2}^-+\Gamma_{\Sigma1}^+\Gamma_{\Sigma2}^-+\Gamma_{\Sigma1}^-\Gamma_{\Sigma2}^+.
\ee
Importantly, it introduces a temperature-dependent prefactor in the expression for the current, see Eq.~\eqref{eq:Jl}:
\be
J_\rmL=\Lambda^{-1}\sum_i\varepsilon_i\left(\Gamma_{\rmL i}^+\Gamma_{\Sigma1}^-\Gamma_{\Sigma2}^--\Gamma_{\rmL i}^-\Gamma_{\Sigma i}^+\Gamma_{\Sigma \bar{i}}^-\right),
\ee
where $\bar{i}=2$ for $i$=1, and viceversa. 

\begin{figure}[t]
\includegraphics[width=\linewidth]{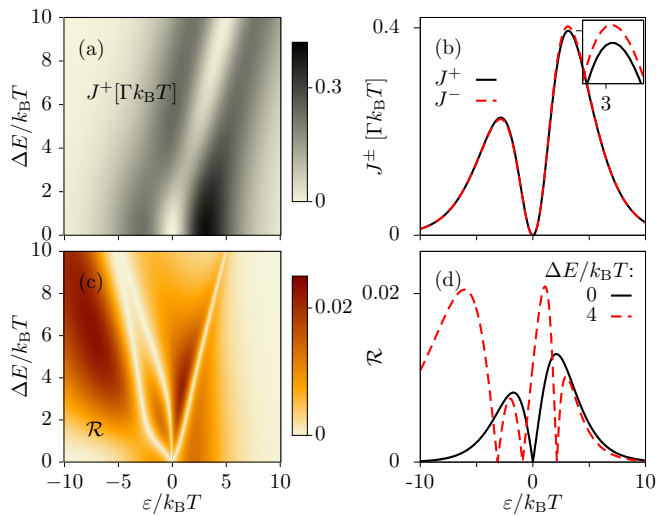}
\caption{Rectification of a single quantum dot whose spin states are split by a magnetic field: $\varepsilon_1=\varepsilon-\Delta E/2$, $\varepsilon_2=\varepsilon+\Delta E/2$, with $\Delta E=g\mu B_{\rm z}$. (a) Heat current for $\Delta T=T/2$, $V=0$, $\Gamma_\rmL=2\Gamma_\rmR=0.2\kBT$, with $\Gamma=\Gamma_\rmL\Gamma_\rmR/\Gamma_\Sigma$. (b) Forward and backward currents for $B_{\rm z}=0$. (c) Rectification coefficient with (d) cuts at zero and finite magnetic field. The inset in (b) zooms in the difference of the two currents around $\varepsilon=3\kBT$. Note that rectification is present even if $B_{\rm z}=0$. }
\label{rectQD}
\end{figure}

With this expression, we can check what the necessary asymmetries are to find a finite rectification. For instance, it is easy to verify that mirror symmetry needs to be broken: if otherwise $\Gamma_{li}=\Gamma_i$, i.e., if tunneling rates only depend on the quantum dot level, we find $J^+=J^-$.

Let us consider the simplest case with energy-independent rates, $\Gamma_{li}=\Gamma_l$, $\forall i$. It is maybe the  most accessible case for experiments.  One way to tune the energy difference of two levels is to introduce a magnetic field $B_{\rm z}$ that induces a Zeeman splitting between the states with opposite z-component, $\Delta E=g\mu_{\rm B}B_{\rm z}$, where $g$ is the gyromagnetic factor, and $\mu_{\rm B}$ is the Bohr magneton. In this case, $\varepsilon_1=\varepsilon-\Delta E/2$, and $\varepsilon_2=\varepsilon+\Delta E/2$. The bare position of the level, $\varepsilon$, can be tuned with a plunger gate. We can clearly distinguish two regimes in Fig.~\ref{rectQD}, depending on whether the Zeeman splitting is smaller or larger than $\kBT$: For $\Delta E<\kBT$, both levels are within the window of thermal excitations and the heat current vanishes close to the symmetric point $\varepsilon=0$. Additionally, at $\varepsilon/\Delta E\approx1/2$ and $-3/2$, we find that $J^+=J^-$. Otherwise, a small but finite rectification appears.

In the regime $\Delta E\gg\kBT$, charge fluctuations affect only one state (at most). When $\varepsilon>0$, the upper state is empty and the system behaves as a single-state quantum dot (discussed at the beginning of this section). In this region, we find a sizable heat current with suppresed rectification (as expected). For $\varepsilon<0$, the lower state is occupied and blocks any transport through the other one, a mechanism related to dynamical channel blockade~\cite{belzig,rf}. It leads to the suppression of transport, so we find a maximal rectification coefficient of tiny heat currents.

Higher rectification coefficients are obtained for the open-circuit case, discussed in Appendix~\ref{app:qd}.

\vspace{-0.4cm}
\subsection{Degenerate levels}
\label{sec_A}
\vspace{-0.3cm}
A particularly interesting case of discussion is when the two states have the same energy, $\varepsilon_1=\varepsilon_2=\varepsilon$. One could na\"ively expect that this configuration presents no rectification, in analogy with the single state case, Eq.~\eqref{onestate}.  However, due to the Coulomb blockade events at different energy levels are correlated, as tunneling into each isolated state is conditioned on the other one being empty. The occupation of each state, in turn, depends on both tunneling transparencies and on temperature, and hence are different depending on which lead is hot, in general.

If the tunneling rates are state-independent, $\Gamma_{li}=\Gamma_l$, this configuration can be mapped to the single-state one by replacing $\Gamma_{l}^+\rightarrow2\Gamma_l^+$. We emphasize that this only affects the tunneling-in rates: while the empty quantum dot has two states that can be occupied, there is only one possible final state when the quantum dot is occupied. The resulting heat current (for $V=0$) reads:
\be
J_{2}=\frac{2\varepsilon\Gamma_\rmL\Gamma_\rmR}{\sum_l\Gamma_l[1+f(\varepsilon,T_l)]}[f(\varepsilon,T_\rmL)-f(\varepsilon,T_\rmR)].
\ee
Note that the denominator of the prefactor now depends on the temperature of the leads. This leads to a finite thermal rectification, which for small temperature gradients can be written as:
\be
J_2^+-J_2^-=\frac{2\Gamma\alpha x^3\kB\Delta T^2}{T\left(3+3\cosh x-\sinh x\right)^2}+{\cal O}\left(\frac{\Delta T}{T}\right)^3,
\ee
with $x=\varepsilon/\kBT$ and $\Gamma=\Gamma_\rmL\Gamma_\rmR/\Gamma_\Sigma$. Note that it relies on a finite tunneling asymmetry $\alpha=(\Gamma_\rmL-\Gamma_\rmR)/\Gamma_\Sigma$. With this result, one immediately finds that the leading order contribution of the rectification coefficient increases linearly with  $\Delta T$:
\be
{\cal R}_2=\left|\frac{\varepsilon\alpha\Delta T}{2\kBT^2\left(3+3\cosh\frac{\varepsilon}{\kBT}-\sinh\frac{\varepsilon}{\kBT}\right)}\right|,
\ee
for small gradients.

\begin{figure}[t]
\includegraphics[width=\linewidth]{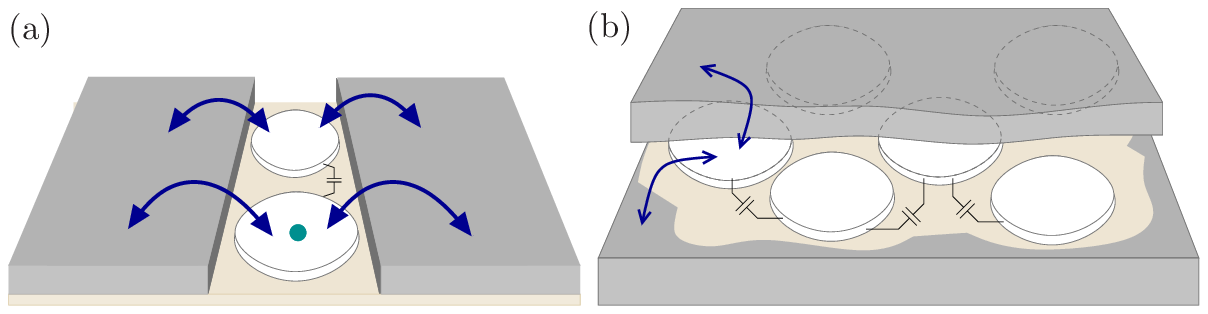}\\
\vspace{0.2cm}
\includegraphics[width=0.8\linewidth]{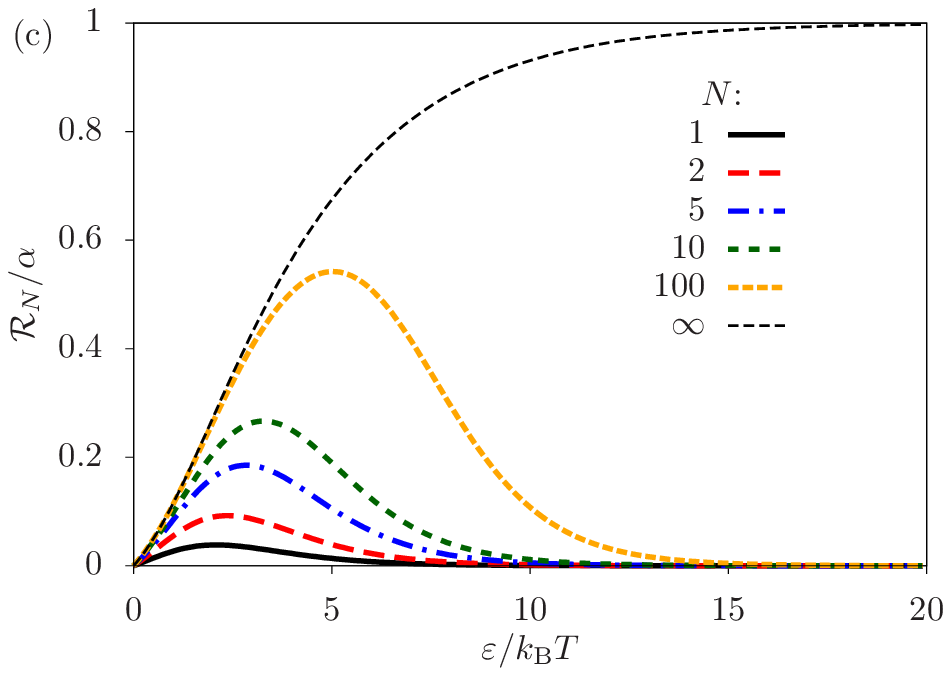}
\caption{Interacting channels in parallel realized in (a) two capacitively coupled quantum dots, and (b) a system of self-assembled quantum dots. (c) Rectification coefficient for a system of $N$ capacitively coupled quantum dots in parallel, with $\Delta T=T/2$.}
\label{rectNQD}
\end{figure}

The effect of the degeneracy of a quantum dot due to spin on the tunneling rates can be explicitly detected in an experiment~\cite{andrea}. It can be modulated by additionally selecting the spin of the injected currents, e.g. with ferromagnetic contacts. A quantum dot coupled to fully-polarized ferromagnetic contacts would recover the results of a single state in Eq.~\eqref{onestate}, see App.~\ref{app:onelevel}. Controlling the polarization of the leads would then switch the rectification on.

\vspace{-0.4cm}
\subsection{Scaling the rectification up}

\vspace{-0.3cm}
This approach opens the possibility of enhancing the rectification by using a larger number of quantum dots.
In this way, the number of accessible states increases, and so does magnitude of the total heat current. Consider for instance an array of $N$ capacitively-coupled quantum dots which are connected to the same two terminals (see Figs.~\ref{rectNQD}(a) and \ref{rectNQD}(b) for possible setups with $N$=2, or larger).  The occupation of one of them increases the charging energy of its neightbours by the Coulomb interaction. Assuming that this  energy is large (compared to the energy of thermal fluctuations, $\kBT_l$), the equations for the new system can also be obtained from the single-state case by replacing $\Gamma_{l}^+\rightarrow2N\Gamma_l^+$ (2 is for spin). This is the opposite case to the one with many non-interacting channels discussed in App.~\ref{app:manyonelevel}.

The rectification in this case reads:
\be
{\cal R}_N=\left|\frac{\alpha(2N{-}1)[f(\varepsilon,T{+}\Delta T){-}f(\varepsilon,T)]}{\left\{2+(2N{-}1)[f(\varepsilon,T{+}\Delta T){+}f(\varepsilon,T)]\right\}}\right|,
\ee
which is plotted in Fig.~\ref{rectNQD}(c). For every given configuration (fixed rates, temperatures, and level position), ${\cal R}_N$ increases with $N$. Of course, a configuration with a large number of quantum dots which are all so strongly coupled that there can only be one electron in all of them is quite unrealistic. It, however, motivates the investigation of thermal effects in densely self-assembled quantum dot layers (which by construction include left-right asymmetries quite naturally) or related multiplexed devices. 

\vspace{-0.5cm}
\section{Coherent tunneling in a double quantum dot}
\label{sec:dqd}

\vspace{-0.3cm}
Some room for improvement is expected for systems with combined mirror asymmetry and energy-resolved tunneling rates. Energy-dependent asymmetries of the tunneling rates in a single quantum dot can be tuned to some extent~\cite{holger}, but they are usually small and difficult to control. To find the desired asymmetry, we consider a double quantum dot coupled in series to the two terminals, as sketched in Fig.~\ref{scheme}(b). Thermoelectric properties of this system have been measured~\cite{chen_thermopower_2000,holgerDQD}. For the sake of simplicity, and in order to isolate the particular effect of this geometry, we will neglect the spin degree of freedom, even when it helps to increase the rectification, as shown in the previous section. The case with spin degeneracy can be recovered again by doubling the tunneling-in rates.

\begin{figure}[t]
\includegraphics[width=0.8\linewidth]{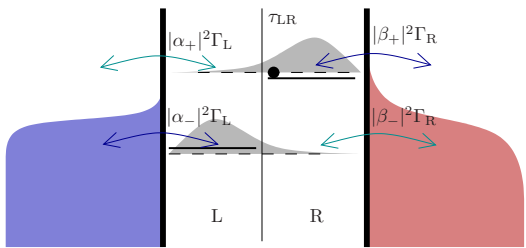}
\caption{Double quantum dot in series. Electron delocalization due to coherent interdot tunneling, $\tau_{\rm LR}$, leads to hybridization of the quantum dot states. The relative weight of the orbitals in each quantum dot affects the tunneling rates.  }
\label{DQDlevels}
\end{figure}

The Hamiltonian of the system takes the form~\cite{vanderwiel_electron_2002}:
\be
\label{hdqd}
\hat H_{\rm DQD}=\sum_{l=\rmL,\rmR}\varepsilon_l\hat n_l-\tau_{\rm LR}\left(\hat c_\rmR^\dagger\hat c_\rmL+{\rm H.c.}\right),
\ee
where $\varepsilon_l$ is the energy of the level of each quantum dot, $l$=L,R, and $\hat n_l$ its occupation operator. Coherent interdot tunneling, $\tau_{\rmL\rmR}$, produces the hybridization of the states $|\rmL\rangle$ and $|\rmR\rangle$, which leads to the formation of molecular-like orbitals: 
\begin{align}
|\pm\rangle = \alpha_\pm |\rmL\rangle -\beta_\pm |\rmR\rangle.
\end{align}
The coefficients $\alpha_\pm=g(2\tau_{\rmL\rmR}/[\varepsilon_\rmL-\varepsilon_\rmR\pm\Delta E])$ and $\beta_\pm=g([\varepsilon_\rmL-\varepsilon_\rmR\pm\Delta E]/2\tau_{\rmL\rmR})$, with $g(x)=(1+x^2)^{-1/2}$ and $\Delta E=\sqrt{(\varepsilon_\rmL-\varepsilon_\rmR)^2+4\tau_{\rmL\rmR}^2}$
come out of the diagonalization of the Hamiltonian~\eqref{hdqd}, also giving the eigenenergies: $E_\pm=(\varepsilon_\rmL+\varepsilon_\rmR\pm\Delta E)/2$. When $\varepsilon_\rmL\neq\varepsilon_\rmR$, the distribution of an electron in one of the eigenstates is not homogeneous for the two dots, as sketched in Fig.~\ref{DQDlevels}. 

In the regime $\tau_{\rmL\rmR} \gg \Gamma$, the dynamics is dominated by the eigenstates. In the basis $|X\rangle=|0\rangle,|\pm\rangle$, the master equation is equivalent to Eq.~\eqref{matrixQD} with $i=\pm$. The tunneling rates from the reservoirs to the eigenstates are determined by the projection of the eigenstates on the localized basis, $\Gamma_{\rmL\pm} = |\alpha_\pm|^2 \Gamma_\rmL$ and $\Gamma_{\rmR\pm} = |\beta_\pm|^2 \Gamma_\rmR$. This way, hybridization effectively introduces mirror-asymmetric and energy-dependent tunneling rates,  even if the barriers are energy-independent and left-right symmetric ($\Gamma_\rmL=\Gamma_\rmR$). Note that, in this case, the rates are symmetric by pairs: $\Gamma_{\rmL\pm}=\Gamma_{\rmR\mp}$. Furthermore, these asymmetries can be experimentally tuned by controlling the splitting $\varepsilon_\rmL-\varepsilon_\rmR$ with gate voltages~\cite{vanderwiel_electron_2002,holgerDQD}.

\begin{figure}
\includegraphics[width=\linewidth]{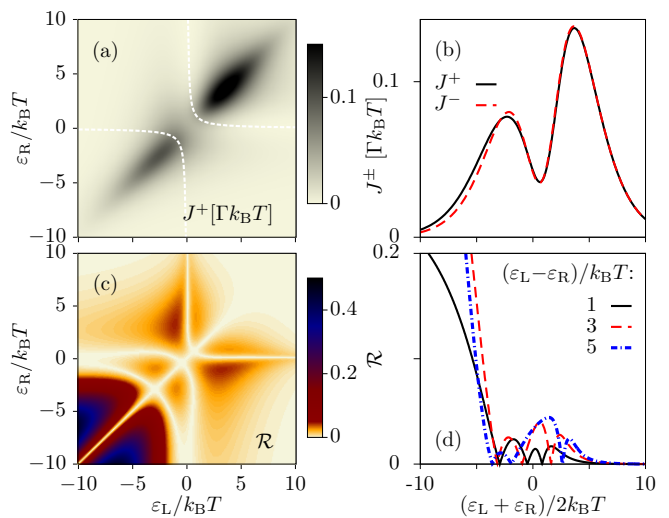}
\caption{Rectification of a double quantum dot whose energy levels, $\varepsilon_\rmL$ and $\varepsilon_\rmR$ are tuned by gate voltages. (a) Heat current for $\Delta T=T/2$, $V=0$, $\tau_{\rmL\rmR}=\kBT$, $\Gamma_\rmL=\Gamma_\rmR=0.2\kBT$, with $\Gamma=\Gamma_\rmL\Gamma_\rmR/\Gamma_\Sigma$. White-dashed lines mark the zeros of the eigenenergies $E_\pm$. (b) Forward and backward currents for $\varepsilon_{\rmR}-\varepsilon_\rmL=\kBT$. (c) Rectification coefficient with (d) cuts at different level splittings. It vanishes when the system is symmetric ($\varepsilon_\rmL=\varepsilon_\rmR$), and close to the conditions $E_\pm\approx0$.}
\label{rectDQD}
\end{figure}

In the limit when the detuning between the two dots is large, each eigenstate recovers the state of a different quantum dot, which is coupled to a single reservoir. If, for instance, $\varepsilon_\rmR\gg\varepsilon_\rmL+\tau_{\rmL\rmR}$, we have $\Gamma_{\rmL-}\approx\Gamma_\rmL$ and $\Gamma_{\rmR+}\approx\Gamma_\rmR$, with vanishing $\Gamma_{\rmL+}$, and $\Gamma_{\rmR-}$. Hence, transport is suppressed, as shown in Fig.~\ref{rectDQD}(a). 

On the other hand, the heat current is maximal around the resonance condition $\varepsilon_\rmL=\varepsilon_\rmR$, where $|\alpha_\pm|=|\beta_\pm|$, cf. Fig.~\ref{rectDQD}. At this condition, the separation of the two levels is minimal and given by $\Delta E=E_+-E_-=2|\tau_{\rmL\rmR}|$. Note also that in this case the system is completely symmetric,
with: $\Gamma_{\rmL\pm}=\Gamma_{\rmR\pm}$, resulting in ${\cal R}=0$.

The current becomes asymmetric as a function of energy due to the Coulomb interaction. The occupation of the lowest energy level suppresses transport through the other, and hence current is reduced when $E_-<0$, see Figs.~\ref{rectDQD}(a) and \ref{rectDQD}(b). If both levels are over the Fermi energy, there is no effective channel blockade. This asymmetry is clearer in the rectification coefficient, which vanishes when the two levels are over the Fermi energy. The rectification rapidly increases when the two orbitals are well below the chemical potential, $E_\pm\ll0$, cf. Figs.~\ref{rectDQD}(c) and \ref{rectDQD}(d). It can in principle be arbitrarily close to the optimal value ${\cal R}=1$. Unfortunately, currents in this region are strongly suppressed and difficult to detect. 

\vspace{-0.4cm}
\subsection{Open circuit configuration}

\vspace{-0.3cm}
The open circuit configuration is interesting by analogy with a purely thermal conductor. Only heat currents flow through the system. The left and right terminals are floating such that a voltage $V_{\rm th}$ develops to satisfy the condition where charge current is zero. This is the thermovoltage appearing in thermoelectric engines~\cite{benenti_fundamental_2017}. It has to be obtained self-consistently for each configuration by solving the equation $I(V_{\rm th})=0$. We assume for simplicity that the voltage is symmetrically developed in the two leads, such that $\mu_\rmL=-eV_{\rm th}/2$ and $\mu_\rmR=eV_{\rm th}/2$.

\begin{figure}[t]
\includegraphics[width=\linewidth]{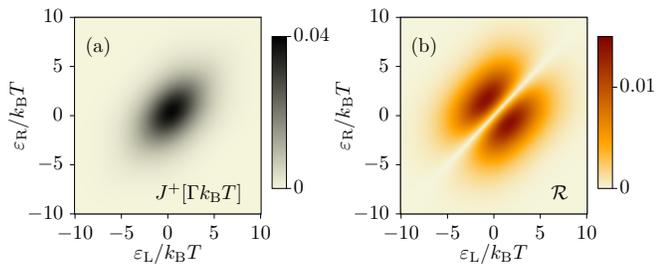}
\caption{Rectification of a double quantum dot in the open-circuit configuration. (a) Heat current and (b) rectification coefficient as functions of the position of the levels, $\varepsilon_1$ and $\varepsilon_2$. The same parameters as in Fig.~\ref{rectDQD} are considered. }
\label{rectDQDnoch}
\end{figure}

In the open circuit configuration, the heat current shows a single peak when both levels are around the Fermi energy, see Fig.~\ref{rectDQDnoch}. The developed voltage suppresses the double-peak structure visible in Fig.~\ref{rectDQD}. Only at $\varepsilon_1=\varepsilon_2=0$, the two orbitals are symmetrically coupled to the leads at $E_\pm=\pm\Delta E/2$, such that $I=0$ and the two cases (open-circuit and short-circuit) coincide. This surprising effect can be understood because when the two channels are over the Fermi energy, only the one with the lowest energy contributes to transport. The system effectively behaves as a single-channel, whose charge and heat currents become proportional. As charge current is zero, also heat vanishes. 

Notably, the rectification is maximal in the region where the heat current peaks, except at the condition $\varepsilon_1=\varepsilon_2$. As discussed above, the tunneling rates are mirror-symmetric at this condition and there is no rectification. 

We emphasize the difference from the behaviour of a two-state quantum dot (as discussed in Sec.~\ref{sec:qd}), discussed in open circuit conditions in the App.~\ref{app:qd}. There, the rectification increases in the region where transport is strongly suppressed.  

\vspace{-0.4cm}
\section{Interference in a triple quantum dot}
\label{sec:tqd}

\vspace{-0.3cm}
In the previous section, we saw that tunneling asymmetries can be introduced via the hybridization of quantum states due to coherent tunneling. However, the largest rectification coefficients occur at conditions where the forward and backward currents are both small. In this section we propose how to enhance these currents further by exploiting 
the effect of coherence, and introducing a setup where left to right trajectories are affected by interference, while right to left ones are not.

This is the case of a triple quantum dot in a triangular geometry, as pictured in Fig.~\ref{scheme}(c). Dots A and B are connected to the left lead and tunnel-coupled to dot C, which is in turn connected to the right lead. The Hamiltonian takes the form,
\be
\label{htqd}
\hat H_{\rm TQD}=\sum_l\varepsilon_l\hat n_l-\sum_{i={\rm A,B}}\tau_{i{\rm C}}\left(\hat c_{\rm C}^\dagger\hat c_i+{\rm H.c.}\right).
\ee
For simplicity, we assume that A and B are only capacitively coupled, $\tau_{\rmA\rmB}=0$, and $\tau_{\rmA\rmC}=\tau_{\rmB\rmC}=\tau$. As we are interested in left-right asymmetries, we will further assume that $\varepsilon_\rmA=\varepsilon_\rmB=\varepsilon_{\rm AB}$, and that the tunneling barriers between the leads and all the three dots are equal.

The eigenstates of this system are:
\begin{align}
|1\rangle&=\left(1+x_+^2\right)^{-1/2}[x_+(|\rmA\rangle+|\rmB\rangle)+|\rmC\rangle]\\
|2\rangle&=2^{-1/2}(|\rmA\rangle-|\rmB\rangle)\\
|3\rangle&=\left(1+x_-^2\right)^{-1/2}[x_-(|\rmA\rangle+|\rmB\rangle)+|\rmC\rangle],
\end{align}
where $x_\pm=(\varepsilon_{\rm AB}-\varepsilon_\rmC\pm\Delta E_{31})/2\tau$ and $\Delta E_{31}=E_3-E_1=\sqrt{(\varepsilon_\rmAB-\varepsilon_\rmC)^2+8\tau^2}$. The eigenenergies read: $E_1=(\varepsilon_\rmAB+\varepsilon_\rmC-\Delta E_{31})/2$, $E_2=\varepsilon_\rmAB$, and $E_3=(\varepsilon_\rmAB+\varepsilon_\rmC+\Delta E_{31})/2$. 

Note that quantum dot C does not contribute to eigenstate $|2\rangle$. Even if A and B are both coupled to C, tunneling is canceled for this particular superposition due to destructive quantum interference. We call it a dark state~\cite{michaelis}, in analogy with quantum optics~\cite{tobias}. This is a crucial point: an electron tunneling from the left lead can in principle enter any of the three eigenstates. Being either in state $|1\rangle$ or in $|3\rangle$, the electron has some finite probability to populate quantum dot C, and therefore to subsequently tunnel out to the right lead and contribute to transport. Differently, if the electron enters state $|2\rangle$, it will block the current (by avoiding any other state to be occupied) until it eventually tunnels back to the left lead. Once this happens (and before it is occupied again), transport is restored.

On the other hand, electrons tunneling from the right lead can only enter two states, $|1\rangle$ or $|3\rangle$, and then there is always a finite probability that it contributes to transport to the left lead. There is no such destructive interference for electrons tunneling from the right. Hence, the asymmetry in the spacial arrangement of quantum dots translates into left and right moving electrons being affected by very different processes.

\begin{figure}[t]
\includegraphics[width=0.78\linewidth]{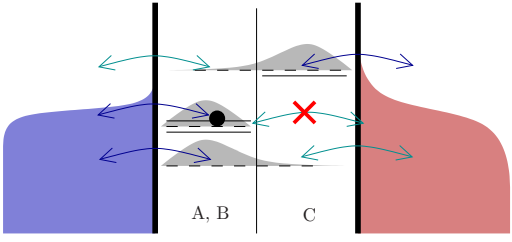}
\caption{Orbitals in a triple quantum dot. When the levels in dots A and B are degenerate, a dark state is formed with forbidden tunneling to the level in dot C. Therefore it is uncoupled from the right lead.}
\label{TQDlevels}
\end{figure}

\begin{figure}[t]
\includegraphics[width=0.95\linewidth]{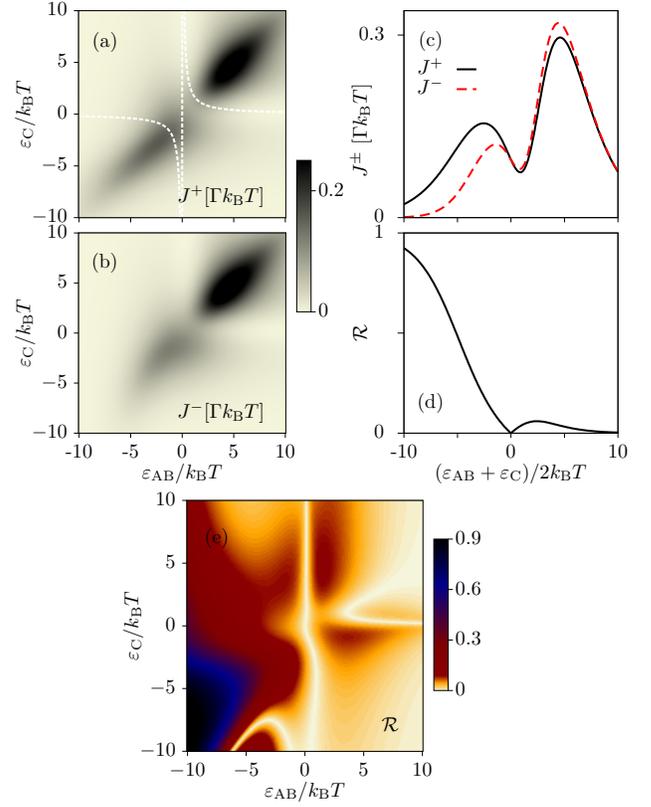}
\caption{Rectification of a triple quantum dot whose energy levels, $\varepsilon_\rmAB$ and $\varepsilon_\rmC$ are tuned by gate voltages. (a) Forward and (b) backward heat currents for $\Delta T{=}T$, $V{=}0$, $\tau{=}\kBT$, $\Gamma_\rmL{=}\Gamma_\rmR{=}0.2\kBT$, with $\Gamma{=}\Gamma_\rmL\Gamma_\rmR/\Gamma_\Sigma$. (c) Forward and backward currents for $\varepsilon_{\rmAB}{-}\varepsilon_\rmC{=}0$. (d) Cut along $\varepsilon_{\rmAB}-\varepsilon_\rmC{=}0$ of the rectification coefficient plotted in (e) for different level positions. When $\varepsilon_\rmAB{<}0$, the occupation of the dark state suppresses the backward current and high rectification coefficients are attained. White-dashed lines in (a) mark the zeros of the eigenenergies $E_i$.}
\label{rectTQD}
\end{figure}

This is reflected in the tunneling rates: $\Gamma_{\rmL i}=(\left|\tilde\alpha_{i\rmA}\right|^2+\left|\tilde\alpha_{i\rmB}\right|^2)\Gamma_\rmL$, and $\Gamma_{\rmR i}=|\tilde\beta_i|^2\Gamma_\rmR$, for $i$=1,3, with $\tilde \alpha_{ij}=\langle i|j\rangle$ and $\tilde \beta_{i}=\langle i|\rmC\rangle$. For the dark state we have $\Gamma_{\rmL2}=\Gamma_\rmL$ and $\Gamma_{\rmR2}=0$. They are illustrated in Fig.~\ref{TQDlevels}. In the limit $\tau\gg\Gamma_l$, the master equation for the states $|X\rangle=|0\rangle,|1\rangle,|2\rangle,|3\rangle$ reads:
\begin{align}
\dot \rho_{ii}&=\Gamma_{\Sigma i}^+\rho_{00}-\Gamma_{\Sigma i}^-\rho_{ii},\quad i{=}1,3\\
\dot \rho_{22}&=\Gamma_{\rmL 2}^+\rho_{00}-\Gamma_{\rmL 2}^-\rho_{22},
\label{matrixTQD}
\end{align}
now taking into account the normalization $\rho_{00}+\rho_{11}+\rho_{22}+\rho_{33}=1$.

The resulting currents are plotted in Fig.~\ref{rectTQD} as the position of the levels $\varepsilon_\rmAB$ and $\varepsilon_\rmC$ are swept. For positive energies, $\varepsilon_\rmAB,\varepsilon_\rmC>0$, small differences between $J_{\rm TQD}^+$ and $J_{\rm TQD}^-$ are mostly attributed to the asymmetries in the tunneling rates due to coherent interdot tunneling, similarly to the effect discussed in Sec.~\ref{sec:dqd}.

Most interestingly, we find a large difference when $\varepsilon_\rmAB<0$. In this region, the dark state is below the chemical potential so it can be populated from the left lead, thus blocking the transport. If $\varepsilon_\rmAB<\kBT_\rmL$, the probability that an electron in state $|2\rangle$ tunnels back to the lead is exponentially suppressed. Hence, the dynamical blockade can not be lifted. Clearly the crossover to this situation occurs at smaller energies (in absolute value) when the left lead is the cold one. This way, the backward current vanishes, while the forward current can still increase due to the onset of transport through state $|3\rangle$.

\begin{figure}[t]
\includegraphics[width=\linewidth]{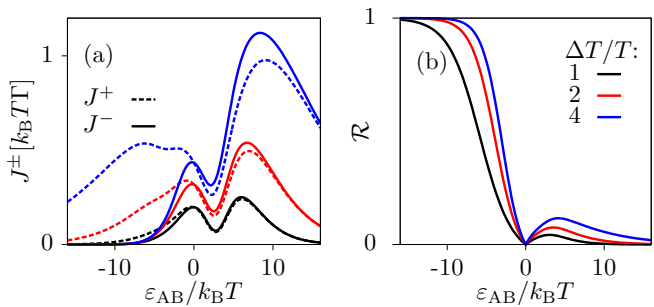}
\caption{Effect of temperature in a triple quantum dot. (a) Heat currents and (b) rectification coefficient for increasing temperature gradients. We assume $\varepsilon_\rmAB=\varepsilon_\rmC$ $V=0$, $\tau=2\kBT$, $\Gamma_\rmL=\Gamma_\rmR=0.04\kBT$, with $\Gamma=\Gamma_\rmL\Gamma_\rmR/\Gamma_\Sigma$. Forward (backward) currents are plotted with dashed (solid) lines. }
\label{rectTQDtemp}
\end{figure}

As the temperature gradient is increased, the contribution of  state $|3\rangle$ to the forward current, $J^+$, increases. It appears as an additional peak when $E_3<0$, see Fig.~\ref{rectTQDtemp}. On the contrary, this signal is not present in the backward current. The onset of the dark state blocking is independent of the temperature of the hot lead and avoids the occupation of $|3\rangle$. This is indeed the desired diode effect: the forward current has a peak where the backward current vanishes. The rectification coefficient is then ${\cal R}\approx1$ for a measurable heat current.

\vspace{-0.4cm}
\section{Coupled quantum dots}
\label{sec:qdqd}

\begin{figure}[t]
\includegraphics[width=0.6\linewidth]{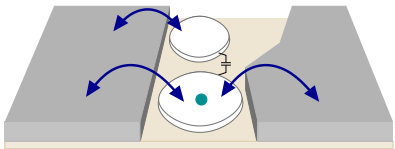}
\caption{Two capacitively-coupled quantum dots,  one of which carries an electron transport, with the other one only supporting fluctuations by being coupled to only one lead. If the capacitive coupling is strong, the occupation of the later acts as a switch by preventing a second electron to tunnel into the conducting dot.} 
\label{schemeQDQD}
\end{figure}

\vspace{-0.3cm}
We can extend the effect shown in the last section, where a high rectification coefficient was produced by the dynamical channel blockade, to get large rectifications for simpler systems.
The strong Coulomb interaction converted the charging/uncharging of the dark state in a switch for the current through the rest of the system. In this section, we present a minimal configuration in which this mechanism is present. It consists of two quantum dots which are capacitively coupled,  as sketched in Fig.~\ref{schemeQDQD}. The coupling is strong enough to avoid two electrons in the system. This system can be realized in semiconductor two-dimensional electron gases~\cite{chan_strongly_2002,hubel_two_2007,hotspots,holger,thgating,hartmann_voltage_2015}, graphene heterostructures~\cite{bischoff_measurement_2015}, metallic islands~\cite{koski_on-chip_2015}, coupled nanowires~\cite{keller_cotunneling_2016}, or corner states in nanowire field-effect transistors~\cite{voisin_few_2014,fernando}. 

We require that one of them is connected to the left and right leads and supports the charge and heat current, whereas the other one is only connected to the left lead. The occupation of the latest dot blocks the current on the former one and therefore works as a switch. Similar processes can be found in single quantum dots with peculiar tunneling couplings~\cite{xue2008thermal}. A related geometry (also with up to one electron) but in a three-terminal configuration has been proposed as a thermal transistor~\cite{thermal,thermal-long} and realized experimentally in metallic Coulomb-blockade islands~\cite{singh_distribution_2016}. 

The advantage of this system is that it is enough that the switch dot is coupled to only one terminal to have the necessary left-right asymmetry. The conducting quantum dot can in principle be totally symmetric. The separation of the conducting and switching states in different quantum dots allows them to be tuned independently. Also, this mechanism does not rely on interference and is hence robust against decoherence and noise sources.

Let us consider spinless electrons, so the rate equations can be obtained as a particular case of the two-state configuration, cf. Eq.~\eqref{matrixQD}, particularized to the case: $\Gamma_{\rm R2}=0$.  In this case, the states $X=1,2$ denote the occupation of the conducting and switch quantum dots, respectively.

The current through the system can be easily obtained, and written in a simple form as:
\be
J_{\rm CQD}=J_1(\varepsilon_1)(1-\rho_{22}),
\ee
in terms of the current through a single channel written in Eq.~\eqref{onestate}. Remarkably, the current is conditioned to the switch dot not being occupied. The steady-state occupation of the latter:
\be
\label{p2}
\rho_{22}=\frac{1-f(\varepsilon_2,T_\rmL)}{1+(\Gamma_{\Sigma1}^+/\Gamma_{\Sigma1})[1-f(\varepsilon_2,T_\rmL)]}
\ee
does not depend on the rate of the switching process, which is therefore determined by the state of the conductor dot, and the position of the level $\varepsilon_2$ with respect to the chemical potential of the left lead.

\begin{figure}[t]
\includegraphics[width=\linewidth]{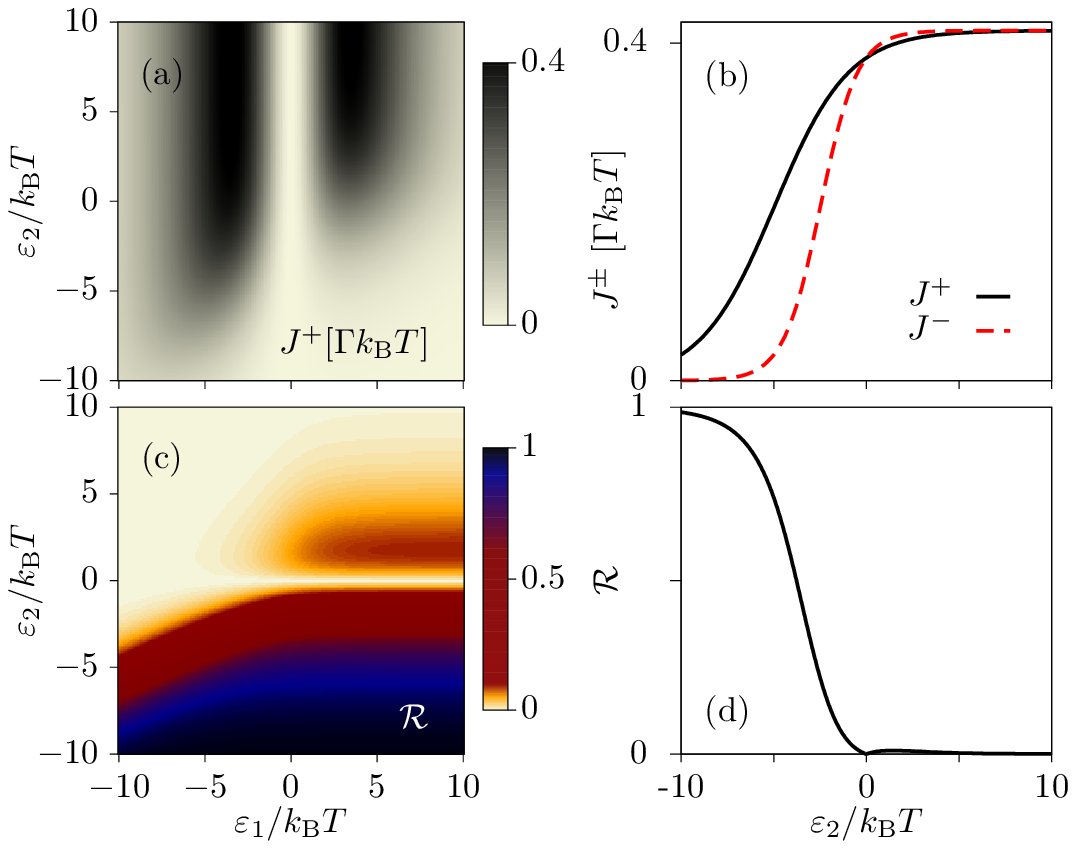}
\caption{Rectification of a system of capacitively quantum dots, one of which is tunneled coupled to one lead, only. (a) Forward heat current for $\Delta T=T$, $V=0$, $\Gamma_{li}=0.2\kBT$, except for $\Gamma_{\rmR 2}=0$, with $\Gamma=\Gamma_\rmL\Gamma_\rmR/\Gamma_\Sigma$. (b) Forward and backward currents along the maximum at $\varepsilon_{1}=-0.36\kBT$. (d) Cut along the same condition of the rectification coefficient plotted in (c) for different level positions. When $\varepsilon_2<0$, the occupation of the coupled dot suppresses the backward current, resulting in high rectification coefficients.}
\label{rectQDQD}
\end{figure}

The cancellation of transport due to the occupation of the second quantum dot can be observed in Fig.~\ref{rectQDQD}. The double peak in the forward heat current vanishes as the energy $\varepsilon_2$ becomes negative. The switch dot is then occupied by an electron, which avoids transitions through the conductor. This effect is most effective when the left lead is cold (i.e. in the backward configuration), because the transition to a state where $\rho_{22}\rightarrow1$ is more pronounced, cf. Fig.~\ref{rectQDQD}(b), following the dependence in the Fermi function~\eqref{p2}. The blockade of the backward current depends exponentially on $\varepsilon_2$, so the rectification coefficient rapidly increases in the region $-\kB(T{+}\Delta T)<\varepsilon_2<-\kBT$ [see Figs.~\ref{rectQDQD}(c) and \ref{rectQDQD}(d)], where $J^+$ is not much affected. 

As all transported electrons have a well defined energy, $\varepsilon_1$, the system does not rectify in the open-circuit configuration, recovering the behaviour of a single-state: since charge and heat currents are proportional to each other, thermal currents vanish at the thermovoltage.

Here we have considered a system of two single-electron quantum dots, whose currents are small and hard to detect. 
However, the same mechanism is in principle applicable to systems that support larger currents (e.g. quantum wires) and are strongly coupled to a switch, opening the way for the definition of thermal diodes which rectify considerably big currents. The switching process can be due to the Coulomb interaction with charges in a quantum dot, as considered here, or due to internal selection rules, e.g. spin blockade in double quantum dots~\cite{ono}.

\vspace{-0.4cm}
\section{Conclusions}
\label{sec:conclusions}

\vspace{-0.3cm}
We have investigated the thermal rectification
of diverse quantum dot systems in the Coulomb blockade regime. Single-electron currents are expected to be small. However, the transport characteristics of these systems can easily be controlled and scaled up to account for greater currents. We also identify  different mechanisms that promote rectification and which can in principle be translated to other systems.

A basic ingredient of our results is the strong Coulomb interaction which introduces correlations between the different conduction channels, although other interactions can also enable the rectification. The details of every configuration determine how the different channels couple to the left and right reservoirs, introducing the necessary asymmetries. 

For a single quantum dot, the presence of two accessible states with broken mirror symmetry is enough to find a finite rectification, even if the two states are degenerate. This can be due, for example, to the spin degree of freedom, in which case the degeneracy can be lifted by means of a magnetic field. Notably, this finding introduces a way to enhance the rectification simply by scaling up the number of interacting channels that contribute to the current. We discuss this possibility by considering a system of several quantum dots coupled in parallel to the two leads.

The asymmetry of the tunneling rates can be additionally controlled in a double quantum dot. Hybridization of the localized states due to coherent tunneling introduces energy-resolved and left-right asymmetric rates that can be tuned by means of gate voltages applied to each quantum dot. As a consequence, larger rectification effects are found, remarkably even approaching ${\cal R}\approx1$. Unfortunately, the huge rectification coefficients correspond to configurations with very small thermal currents. 

Considering that the conducting channels interact with an energy level that is only coupled to one of the leads,
the occupation of this level will act as a switch. 
This way, the mirror symmetry of the conducting channels is broken, and the current strongly depends on the switch state being below the chemical potential of its lead. 
The blockade is lifted by thermal fluctuations, which introduces a temperature-dependent threshold. 
This introduces a huge rectification effect as the presence of a current relies on whether the switch is coupled to the hot or to the cold terminal.

We use this effect in two different configurations: In a triangular triple quantum dot, tunneling interferences lead to the formation of a transport dark state in the two leftmost quantum dots, which avoids tunneling to the right one. An exponential suppression of the backward current is found for configurations where the forward one shows a resonance.

In a system of capacitively-coupled quantum dots, one of them serves as a conductor, while the other one is tunnel-coupled to one lead, only. The rectification coefficient can in this case be controlled with a single gate voltage coupled to the single-terminal quantum dot. This configuration is of experimental relevance~\cite{holger,chan_strongly_2002,hubel_two_2007,thgating,hartmann_voltage_2015,koski_on-chip_2015,bischoff_measurement_2015,keller_cotunneling_2016,singh_distribution_2016} and can readily be tested. 

For typical experimental conditions in semiconductor quantum dots with tunneling rates $\Gamma\sim10$~GHz and $T\sim100$~mK, heat currents would be of the order of 1~{\si fW}, well within present day resolution~\cite{dutta_thermal_2017}. In quantum dots defined in two dimensional electron gases, the regime of application of our results is restricted to low temperatures (of the order of 0.1--1~K) where Coulomb blockade effects have been observed. The application of large temperature gradients is an issue in two dimensional electron gases, but some advances in quantum dots embedded into nanowires has been recently achieved~\cite{josefsson_quantum_2017}. Also, recent room temperature detection of Coulomb blockade in nanoparticles~\cite{zheng_room_2015}, and quantum inteference in molecular junctions~\cite{fracasso_evidence_2011,taniguchi_dependence_2011,guedon_observation_2012,arroyo_signatures_2013,liu_gating_2016} are promising advances toward the application of the effects discussed here in thermal devices. 

Here we have restricted ourselves to the weak-coupling regime. Exploring these interacting effects in full coherent transport and accounting for the effect of possible sources of dephasing~\cite{dani} remain as issues for future work.

\vspace{-0.3cm}
\acknowledgments
\vspace{-0.3cm}
We enjoyed discussions with Fernando Gonz\'alez-Zalba, Milena Grifoni, Clive Emary, Daniel Manzano, David S\'anchez, and thank Holger Thierschmann for useful comments on the manuscript. We acknowledge financial support from the Spanish MINECO via grants No. FIS2015-74472-JIN (AEI/FEDER/UE), No. MTM2017-84446-C2-2-R and No. MTM2014-56948-C2-2-P, and the Ram\'on y Cajal program RYC-2016-20778. M.R.G. also acknowledges support from MECD through the FPU program.

\appendix

\vspace{-0.3cm}
\section{Single non-degenerate level model}
\label{app:onelevel}

\vspace{-0.3cm}
Let us consider the well known model of a quantum dot with a single non-degenerate level at energy $\varepsilon_1$. It is usually used when the spin degree of freedom does not play any role. The two states of the system are defined by whether it is empty or occupied: $|X\rangle=|0\rangle,|1\rangle$. The rate equation simply reads:
\be
\dot\rho_{11}=\Gamma_{\Sigma1}^+\rho_{00}-\Gamma_{\Sigma1}^-\rho_{11},
\ee
with $\Gamma_{\Sigma1}^\pm=\Gamma_{\rmL1}^\pm+\Gamma_{\rmR1}^\pm$. In the stationary limit, it is easy to obtain the steady state occupation: $\rho_{00}=\Gamma_{\Sigma1}^-/(\Gamma_{\Sigma1}^++\Gamma_{\Sigma1}^-)$, and $\rho_{11}=\Gamma_{\Sigma1}^+/(\Gamma_{\Sigma1}^++\Gamma_{\Sigma1}^-)$. The denominator warranties the conservation of probability, $1=\rho_{00}+\rho_{11}$. It is independent of the lead temperature, as $\Gamma_{l1}^++\Gamma_{l1}^-=\Gamma_{l1}$.
With these, it is immediate to obtain the charge current: 
\be
\label{curronelev}
I_1(\varepsilon)=e\frac{\Gamma_{\rmL1}\Gamma_{\rmR1}}{\Gamma_{\rmL1}{+}\Gamma_{\rmR1}}[f(\varepsilon,T_\rmL)-f(\varepsilon,T_\rmR)].
\ee
The heat current is tightly coupled: $J_1/I_1=\varepsilon_1/e$, resulting in Eq.~\eqref{onestate}. 

The currents in Eqs.~\eqref{curronelev} and \eqref{onestate} only depend on temperature through the difference of Fermi functions. Hence, exchanging them results in a global change of sign, i.e. there is no rectification in short-circuit even when the tunneling rates ($\Gamma_{\rmR1}$ and $\Gamma_{\rmL1}$) were not equal. Note that in the case of asymmetric tunneling rates the state occupation, $\rho_{11}$, does change when the temperature difference is reversed. This effect, combined with interaction between levels is what enables rectification in Sec. \ref{sec:qd}.

The tight-coupling relation also avoids rectification in open-circuit, as it involves $I_1=0$.

\vspace{-0.3cm}
\section{Many single non-degenerate level model}
\label{app:manyonelevel}
\vspace{-0.3cm}
The argument in App.~\ref{app:onelevel} holds if one has several copies of the same system which do not interact to each other:
\be
J=\sum_i J_i=\sum_i\varepsilon_i\frac{\Gamma_{\rmL i}\Gamma_{\rmR i}}{\Gamma_{\rmL i}{+}\Gamma_{\rmR i}}[f(\varepsilon_i,T_\rmL)-f(\varepsilon_i,T_\rmR)].
\ee 
The prefactor in the previous expression depends only on the couplings of each channel, and is temperature-independent, again resulting in ${\cal R}=0$ for the short-circuit configuration. 

The open circuit is in this case different: if not all channels have the same energy, tight charge-energy coupling does not hold and therefore rectification is finite.

\vspace{-0.3cm}
\section{Quantum dot in open circuit}
\label{app:qd}

\vspace{-0.3cm}
We consider the two-state quantum dot discussed in Sec.~\ref{sec:qd} under the open-circuit conditions. 

\begin{figure}[b]
\includegraphics[width=\linewidth]{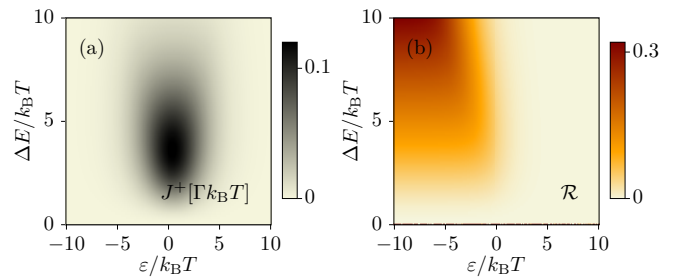}
\caption{Rectification of a two-state quantum dot in open-circuit. (a) Heat current and (b) rectification coefficient as functions of the position of the level and an applied magnetic field. The same parameters as in Fig.~\ref{rectQD} are considered. }
\label{rectQDnoch}
\end{figure}

The heat current characteristics are strongly affected, as shown in Fig.~\ref{rectQDnoch}. First of all, there is no heat flow whatsoever for the condition $\Delta E=0$: since the two states have the same energy, charge and heat currents become proportional to each other, 
\be
J_{2,\rm o-c}=\left(\frac{\varepsilon}{e}+\frac{V}{2}\right) I_{2,\rm o-c}. 
\ee
This is the so-called tight-coupling limit. It follows trivially that the heat current will vanish as well.

The tight coupling is lifted under a finite level splitting, $\Delta E\neq0$, e.g. again due to an applied magnetic field. The cancellation of the charge current does no longer imply that heat flows vanish. Indeed, heat unavoidably flows from the hot to the cold terminal, showing a single peak structure confined in the region where $\varepsilon_2>0$ and $\varepsilon_1<0$. For positive energies, $\varepsilon_2>\varepsilon_1>0$, the upper level is rarely occupied, so the system behaves a single level showing no current in open-circuit. When both levels are negative, the rectification increases linearly. A compromise between large rectification and non-vanishing currents is found then for $\varepsilon_2\approx0$ and $\Delta E\approx3\kBT$.

\end{document}